\input amstex
\documentstyle{amsppt}
%
\catcode`@=11
\redefine\output@{%
  \def\break{\penalty-\@M}\let\par\endgraf
  \ifodd\pageno\global\hoffset=105pt\else\global\hoffset=8pt\fi  
  \shipout\vbox{%
    \ifplain@
      \let\makeheadline\relax \let\makefootline\relax
    \else
      \iffirstpage@ \global\firstpage@false
        \let\rightheadline\frheadline
        \let\leftheadline\flheadline
      \else
        \ifrunheads@ 
        \else \let\makeheadline\relax
        \fi
      \fi
    \fi
    \makeheadline \pagebody \makefootline}%
  \advancepageno \ifnum\outputpenalty>-\@MM\else\dosupereject\fi
}
\def\Beta{\mathchar"0\hexnumber@\rmfam 42}
\redefine\mm@{2010} 
\catcode`\@=\active
\nopagenumbers

\def\grad{\operatorname{grad}}

\def\negskp{\hskip -2pt}

\gdef\blue#1{#1}

\gdef\darkred#1{#1}
\catcode`#=11\def\diez{#}\catcode`#=6
\catcode`&=11\catcode`&=4
\catcode`_=11\def\podcherkivanie{_}\catcode`_=8
\catcode`\^=11\catcode`\^=7
\catcode`~=11\catcode`~=\active
\def\mycite#1{\cite{\blue{#1}}\immediate\special{ps:
     ShrHPSdict begin /ShrBORDERthickness 0 def}}
\def\myciterange#1#2#3#4{\cite{\blue{#2#3#4}}\immediate\special{ps:
     ShrHPSdict begin /ShrBORDERthickness 0 def}}
\def\mytag#1{%
    \tag#1}
\def\mythetag#1{\thetag{\blue{#1}}\immediate\special{ps:
     ShrHPSdict begin /ShrBORDERthickness 0 def}}
\def\myrefno#1{\no#1}
\def\myhref#1#2{\blue{#2}\immediate\special{ps:
     ShrHPSdict begin /ShrBORDERthickness 0 def}}

\def\mytheorem#1{\csname proclaim\endcsname{Theorem #1}}
\def\mytheoremwithtitle#1#2{\csname proclaim\endcsname{Theorem #1#2}}

\def\mylemma#1{\csname proclaim\endcsname{Lemma #1}}
\def\mylemmawithtitle#1#2{\csname proclaim\endcsname{Lemma #1#2}}
\def\mythelemma#1{\blue{#1}\immediate\special{ps:
     ShrHPSdict begin /ShrBORDERthickness 0 def}}
\def\mycorollary#1{\csname proclaim\endcsname{Corollary #1}}

\def\mydefinition#1{\definition{Definition #1}}

\def\myconjecture#1{\csname proclaim\endcsname{Conjecture #1}}
\def\myconjecturewithtitle#1#2{\csname proclaim\endcsname{Conjecture #1#2}}

\def\myproblem#1{\csname proclaim\endcsname{Problem #1}}
\def\myproblemwithtitle#1#2{\csname proclaim\endcsname{Problem #1#2}}

\pagewidth{360pt}
\pageheight{606pt}
\topmatter
\title
On cylindrical regression in three-dimensional Euclidean space.
\endtitle
\author
O. V. Ageev, R. A. Sharipov
\endauthor
\address Self-employed individual, Ufa, Russia
\endaddress
\email ageev-ufa\@yandex.ru
\endemail
\address Bashkir State University, 32 Zaki Validi street, 450074 Ufa, Russia
\endaddress
\email r-sharipov\@mail.ru
\endemail
\urladdr
\vtop to 20pt{\hsize=280pt\noindent
\myhref{http://ruslan-sharipov.ucoz.com}
{http:/\negskp/ruslan-sharipov.ucoz.com}\newline
\myhref{http://freetextbooks.narod.ru}
{http:/\negskp/freetextbooks.narod.ru}\vss}
\endurladdr
\abstract
    The three-dimensional cylindrical regression problem is a problem of finding 
a cylinder best fitting a group of points in three-dimensional Euclidean space. 
The words best fitting are usually understood in the sense of the minimum root mean 
square deflection of the given points from a cylinder to be found. In this form the 
problem has no analytic solution. If one replaces the root mean square averaging
by a certain biquadratic averaging, the resulting problem has an almost analytic 
solution. This solution is reproduced in the present paper in a coordinate-free form.
\endabstract
\endtopmatter
\TagsOnRight
\document

\rightheadtext{On cylindrical regression \dots}
\head
1. Introduction.
\endhead
\parshape 11 0pt 360pt 0pt 360pt 0pt 360pt 0pt 360pt 0pt 360pt 0pt 360pt 
0pt 360pt 0pt 360pt 0pt 360pt 0pt 360pt 180pt 180pt
    Linear, circular, elliptic, and ellipsoidal regression problems are presented in many 
sources (see \myciterange{1}{1}{--\kern 1pt}{6}). Cylinder is one more geometric shape
commonly used in machine design and in technical drawing. The cylindrical regression problem 
is also presented in many sources (see \myciterange{7}{7}{--\kern 1pt}{11}). However, its
solution is usually given in the form of a computational algorithm using iterative approximations. 
The exception is \mycite{12}. In section 7 of \mycite{12} the almost analytic 
solution of the problem is found. Unfortunately this solution is expressed in a 
semi-coordinate form associating an orthonormal triple of vectors $\bold U$, $\bold V$, 
$\bold W$ with the cylinder being considered. Moreover it uses the concept of the center 
point $\bold C$, which is ambiguous for a cylinder. \vadjust{\vskip 5pt\hbox 
to 0pt{\kern 15pt \includegraphics{cylinder.eps}\hss}\vskip -5pt}Our goal in 
this paper is to reproduce the almost analytic solution of the cylindrical regression problem 
from \mycite{12} in a coordinate-free form thus making it more clear and reader-friendly.
\par
\head
\parshape 1 180pt 180pt
2. A cylinder and its axis.
\endhead
\parshape 10 180pt 180pt 180pt 180pt 180pt 180pt 180pt 180pt
180pt 180pt 180pt 180pt 180pt 180pt 180pt 180pt 180pt 180pt 
0pt 360pt  
     Any cylinder is given by its radius $\rho$ and its axis (see Fig\.~2.1). The axis of a 
cylinder is a straight line. Usually a straight line is given by the equation 
$$
\hskip -2em
\bold r=\bold r_0+\bold a\,t,
\mytag{2.1}
$$
where $\bold r_0$ is the radius-vector of some fixed point $A$ of the line, $\bold a$ is 
some non-zero vector on the line, and $t$ is a scalar parameter. In \mycite{12} the point
$A$ is denoted through $\bold C$ and is called the center of a cylinder. As it was noted in 
\mycite{6} the choice of such a point is not unique. In order to avoid the ambiguity in the
choice of the initial point $A$ on the line (see Fig\.~2.1) the parametric equation
\mythetag{2.1} was replaced by the following vectorial non-parametric equation: 
$$
\hskip -2em
[\bold r,\bold a]=\bold b\text{, \ where \ }\bold b\perp\bold a
\mytag{2.2}
$$
(see \mycite{6} and \mycite{13}). The square brackets in \mythetag{2.2} stand for 
the vector product\footnotemark\ operation. \footnotetext{\ It is also called the cross
product, i\.\,e\. $[\bold x,\bold y]=\bold x\times\bold y$.}\ The vector $\bold b$ is 
produced from $\bold r_0$ by means of the formula
$$
\hskip -2em
\bold b=[\bold r_0,\bold a],
\mytag{2.3}
$$
however it has no ambiguity since the right hand side of \mythetag{2.3} is invariant 
with respect to the transformation $\bold r_0\to\bold r_0 + \bold a\,t$.\par
\head
3. The statement of the problem.
\endhead
     Let $X_1,\,\ldots,\,X_n$ be a group of points in the space given by their
radius-vectors $\bold r_1,\,\ldots,\,\bold r_n$. The cylindrical regression problem
in our case consists in finding three parameters --- two vectors $\bold a$ and 
$\bold b$ and one scalar $\rho$ that determine the radius and the axis of a cylinder 
best fitting the group of points $X_1,\,\ldots,\,X_n$ in the sense of the following
biquadratically averaged deflection 
$$
\hskip -2em
\bar D^{\kern 1.3pt 2}=\frac{1}{n}\sum^n_{i=1}d_i^{\kern 1.3pt 2}(2\,\rho\pm d_i)^2.
\mytag{3.1}
$$ 
Here $d_1,\,\ldots,\,d_n$ are the distances from the points $X_1,\,\ldots,\,X_n$ to 
the surface of a cylinder to be found. The sign plus in \mythetag{3.1} is taken for 
exterior points and minus for interior points. In \mycite{12} the quantity 
\mythetag{3.1} is called the least-squares error function\footnotemark. It differs from the 
standard least-squares sum
\footnotetext{\ Up to the constant factor $1/n$.}
\adjustfootnotemark{-2}
$$
\hskip -2em
\bar d^{\kern 1.3pt 2}=\frac{1}{n}\sum^n_{i=1}d_i^{\kern 1.3pt 2},
\mytag{3.2}
$$
but coincides with the sum \thetag{3.3} used in \mycite{3} in the case of a circle. 
The sum \mythetag{3.1} is equivalent to the sum \mythetag{3.2} in the sense that 
both $\bar D$ and $\bar d$ tend to zero as $d_i\to 0$. In this case 
$\bar D/\bar d\to 2\,\rho$.\par
\head
4. Initial steps for solving the problem.
\endhead
     The distance from the point $X_i$ to the line \mythetag{2.1}, which is the axis
of the cylinder (see Fig\.~2.1), is given by the following formula 
$$
\hskip -2em
\rho_i=\frac{|[\bold r_i-\bold r_0,\bold a]|}{|\bold a|}. 
\mytag{4.1}
$$
Without loss of generality we can assume that
$$
\hskip -2em
|\bold a|=1. 
\mytag{4.2}
$$
Then, taking into account \mythetag{2.3} and \mythetag{4.2}, from
\mythetag{4.1} we derive
$$
\hskip -2em
\rho_i=|[\bold r_i,\bold a]-\bold b|. 
\mytag{4.3}
$$
\par
     The distance from the point $X_i$ to the surface of the cylinder 
(see Fig\.~2.1) is obviously given by the following formula:
$$
\hskip -2em
d_i=|\rho_i-\rho|.
\mytag{4.4}
$$
Let's apply \mythetag{4.4} to \mythetag{3.1}, As a result we obtain
$$
\hskip -2em
\bar D^{\kern 1.3pt 2}=\frac{1}{n}\sum^n_{i=1}(\rho_i-\rho)^2(\rho_i+\rho)^2
=\frac{1}{n}\sum^n_{i=1}(\rho_i^{\kern 1.3pt 2}-\rho^{\kern 1pt 2})^2.
\mytag{4.5}
$$ 
\mydefinition{4.1} A cylinder with the radius $\rho$ and the axis given by the equation 
\mythetag{2.3}, where $|\bold a|=1$, is called an {\it optimal cylinder\/} best fitting the 
points $X_1,\,\ldots,\,X_n$ if the quantity \mythetag{4.5} takes its minimal value.
\enddefinition
     Expanding \mythetag{4.5}, we derive the following expression for 
$\bar D^{\kern 1.3pt 2}$:
$$
\hskip -2em
\bar D^{\kern 1.3pt 2}=\frac{1}{n}\sum^n_{i=1}\rho_i^{\kern 1.3pt 4}
-\frac{2\,\rho^{\kern 1pt 2}}{n}\sum^n_{i=1}\rho_i^{\kern 1.3pt 2}
+\rho^{\kern 1pt 4}.
\mytag{4.6}
$$ 
The expression in the right hand side of \mythetag{4.6} is biquadratic with respect
to the variable $\rho$. It takes its minimal value if $\rho$ is given by the
formula
$$
\hskip -2em
\rho^{\kern 1pt 2}=\frac{1}{n}\sum^n_{i=1}\rho_i^{\kern 1.3pt 2}.
\mytag{4.7}
$$
The formula \mythetag{4.7} coincides with the formula \thetag{117} in \mycite{12}. 
Substituting \mythetag{4.7} back into \mythetag{4.6}, we derive the following 
formula:
$$
\hskip -2em
\bar D^{\kern 1.3pt 2}=\frac{1}{n}\sum^n_{i=1}\rho_i^{\kern 1.3pt 4}
-\biggl(\frac{1}{n}\sum^n_{i=1}\rho_i^{\kern 1.3pt 2}\biggr)^{\!\lower 2pt\hbox{$\ssize 2$}}.
\mytag{4.8}
$$ 
The formula \mythetag{4.8} is similar to the formula \thetag{3.6} in \mycite{3}. The 
formula \mythetag{4.3} yields the following expression for $\rho_i^{\kern 1.3pt 2}$ in
\mythetag{4.8}: 
$$
\hskip -2em
\rho_i^{\kern 1.3pt 2}=|\bold b|^2 
-2\,([\bold r_i,\bold a],\bold b)
+|[\bold r_i,\bold a]|^2. 
\mytag{4.9}
$$
The round brackets in \mythetag{4.9} denote the scalar product\footnotemark\ 
operation.\par
\footnotetext{\ It is also called the dot product, i\.\,e\. $(\bold x,\bold y)=\bold x\cdot
\bold y$.}\adjustfootnotemark{-1}
     Now let's substitute \mythetag{4.9} back into \mythetag{4.8}. This yields
$$
\hskip -2em
\gathered
\bar D^{\kern 1.3pt 2}=\frac{1}{n}\sum^n_{i=1}\Bigl(|\bold b|^2 
-2\,([\bold r_i,\bold a],\bold b)
+|[\bold r_i,\bold a]|^2\Bigr)^{\!\lower 2pt\hbox{$\ssize 2$}}\,-\\
-\biggl(\frac{1}{n}\sum^n_{i=1}\Bigl(|\bold b|^2 
-2\,([\bold r_i,\bold a],\bold b)+|[\bold r_i,\bold a]|^2\Bigr)
\biggr)^{\!\lower 2pt\hbox{$\ssize 2$}}.
\endgathered
\mytag{4.10}
$$ 
Upon expanding \mythetag{4.10} we find that the fourth order terms and the third
order terms with respect to $\bold b$ do cancel each other. As a result we get
$$
\gathered
\bar D^{\kern 1.3pt 2}=\frac{4}{n}\sum^n_{i=1}([\bold r_i,\bold a],\bold b)^2
-4\,\biggl(\frac{1}{n}\sum^n_{i=1}([\bold r_i,\bold a],\bold b)
\biggr)^{\!\lower 2pt\hbox{$\ssize 2$}}\,+\\
+\,4\,\biggl(\frac{1}{n}\sum^n_{i=1}([\bold r_i,\bold a],\bold b)\biggr)
\biggl(\frac{1}{n}\sum^n_{i=1}|[\bold r_i,\bold a]|^2\biggr)
-\frac{4}{n}\sum^n_{i=1}([\bold r_i,\bold a],\bold b)
\,|[\bold r_i,\bold a]|^2\,+\\
+\,\frac{1}{n}\sum^n_{i=1}|[\bold r_i,\bold a]|^4
-\biggl(\frac{1}{n}\sum^n_{i=1}|[\bold r_i,\bold a]|^2
\biggr)^{\!\lower 2pt\hbox{$\ssize 2$}}.
\endgathered
\quad
\mytag{4.11}
$$\par      
     In \mythetag{4.11} we see the following combination of the vectorial and scalar 
products: $([\bold r_i,\bold a],\bold b)$. Such a combination is known as the
mixed product (see \mycite{13}):
$$
\hskip -2em
([\bold r_i,\bold a],\bold b)=(\bold b,[\bold r_i,\bold a])=(\bold b,\bold r_i,\bold a).
\mytag{4.12}
$$
The mixed product $(\bold b,\bold r_i,\bold a)$ is completely skew-symmetric, i\.\,e\.
it is skew-symmetric with respect to any pair of its multiplicands. Therefore
$$
\hskip -2em
(\bold b,\bold r_i,\bold a)=-(\bold r_i,\bold b,\bold a)=(\bold r_i,\bold a,\bold b)
=(\bold r_i,[\bold a,\bold b]).
\mytag{4.13}
$$
Combining \mythetag{4.12} and \mythetag{4.13}, we derive 
$$
\hskip -2em
([\bold r_i,\bold a],\bold b)=(\bold r_i,[\bold a,\bold b]).
\mytag{4.14}
$$
Relying on \mythetag{4.14}, we introduce the following vector $\bold c$:
$$
\hskip -2em
\bold c=[\bold a,\bold b].
\mytag{4.15}
$$
Since $\bold b\perp\bold a$ (see \mythetag{2.2}), there is an inverse formula 
expressing $\bold b$ through $\bold c$:
$$
\hskip -2em
\bold b=-\frac{[\bold a,\bold c]}{|\bold a|^2}.
\mytag{4.16}
$$\par
     Let's apply \mythetag{4.15} to \mythetag{4.14} and then substitute 
\mythetag{4.14} back into \mythetag{4.11}. As a result we derive the following
formula for $\bar D^{\kern 1.3pt 2}$:
$$
\gathered
\bar D^{\kern 1.3pt 2}=\frac{4}{n}\sum^n_{i=1}(\bold r_i,\bold c)^2
-4\,\biggl(\frac{1}{n}\sum^n_{i=1}(\bold r_i,\bold c)
\biggr)^{\!\lower 2pt\hbox{$\ssize 2$}}\,+\\
+\,4\,\biggl(\frac{1}{n}\sum^n_{i=1}(\bold r_i,\bold c)\biggr)
\biggl(\frac{1}{n}\sum^n_{i=1}|[\bold r_i,\bold a]|^2\biggr)
-\frac{4}{n}\sum^n_{i=1}(\bold r_i,\bold c)
\,|[\bold r_i,\bold a]|^2\,+\\
+\,\frac{1}{n}\sum^n_{i=1}|[\bold r_i,\bold a]|^4
-\biggl(\frac{1}{n}\sum^n_{i=1}|[\bold r_i,\bold a]|^2
\biggr)^{\!\lower 2pt\hbox{$\ssize 2$}}.
\endgathered
\quad
\mytag{4.17}
$$
The first two terms in \mythetag{4.17} are quadratic with respect to the vector 
$\bold c$. They can be written as $4\,Q(\bold c,\bold c)$, where $Q$ is a
quadratic form:
$$
\hskip -2em
Q(\bold c,\bold c)=\frac{1}{n}\sum^n_{i=1}(\bold r_i,\bold c)^2
-\biggl(\frac{1}{n}\sum^n_{i=1}(\bold r_i,\bold c)
\biggr)^{\!\lower 2pt\hbox{$\ssize 2$}}.
\mytag{4.18}
$$
The next two terms in \mythetag{4.17} are linear with respect to the vector 
$\bold c$. They can be written as $-4\,(\bold L,\bold c)$, where $\bold L$ is
the following vector:
$$
\hskip -2em
\bold L=\frac{1}{n}\sum^n_{i=1}\bold r_i\,|[\bold r_i,\bold a]|^2
-\biggl(\frac{1}{n}\sum^n_{i=1}\bold r_i\biggr)
\biggl(\frac{1}{n}\sum^n_{i=1}|[\bold r_i,\bold a]|^2\biggr).
\mytag{4.19}
$$
The last two terms in \mythetag{4.17} do not depend on $\bold c$. They can
be written as 
$$
\hskip -2em
M=\frac{1}{n}\sum^n_{i=1}|[\bold r_i,\bold a]|^4
-\biggl(\frac{1}{n}\sum^n_{i=1}|[\bold r_i,\bold a]|^2
\biggr)^{\!\lower 2pt\hbox{$\ssize 2$}}.
\mytag{4.20}
$$
Combining \mythetag{4.18}, \mythetag{4.19}, and \mythetag{4.20}, we can write
\mythetag{4.17} as 
$$
\hskip -2em
\bar D^{\kern 1.3pt 2}=4\,Q(\bold c,\bold c)-4\,(\bold L,\bold c)+M.
\mytag{4.21}
$$
The last term $M$ in \mythetag{4.21} is a constant with respect to the variable 
$\bold c$, though it depends on the variable $\bold a$. Similarly $\bold L$
is a constant vector with respect to the variable $\bold c$, though it depends on 
the variable $\bold a$. The quadratic form $Q(\bold c,\bold c)$ does not depend
on $\bold a$.\par
     Note that the formula \mythetag{4.21} is similar to the formula \thetag{3.7}
in \mycite{3}. The quadratic form $Q(\bold c,\bold c)$ in it coincides with the
quadratic form $Q(\bold n,\bold n)$ given by the formula \thetag{2.19} in 
\mycite{3}. It does actually coincides with the form $Q(\bold a,\bold a)$ given
by the formula \thetag{4.7} in \mycite{6}, though the formula \thetag{4.7} in 
\mycite{6} looks somewhat different from \mythetag{4.18}. The form $Q(\bold c,\bold c)$
is called the {\it non-flatness form} for a group of points in \mycite{3}. It is called
the {\it non-linearity form} in \mycite{6}. Both terms are consistent regarding the
applications of the form $Q$ in \mycite{3} and \mycite{6}.\par 
\mylemma{4.1} The expression for $Q(\bold c,\bold c)$ in \mythetag{4.18} is invariant
with respect to the transformation $\bold r_i\to\bold r_i-\bold p$, where $\bold p$ is an
arbitrary constant vector. 
\endproclaim
     The proof is pure calculations upon substituting $\bold r_i-\bold p$ for 
$\bold r_i$ into \mythetag{4.18}.\par
     Now, if we define the center of mass for the group of points $X_1,\,\ldots,\,X_n$ 
by means of the formula for its radius-vector
$$
\hskip -2em
\bold r_{\text{cm}}=\frac{1}{n}\sum^n_{i=1}\bold r_i,
\mytag{4.22}
$$
then we can choose $\bold p=\bold r_{\text{cm}}$ and apply Lemma~\mythelemma{4.1}. As 
a result we obtain the following formula for the quadratic form $Q(\bold c,\bold c)$:  
$$
\hskip -2em
Q(\bold c,\bold c)=\frac{1}{n}\sum^n_{i=1}(\bold r_i-\bold r_{\text{cm}},\bold c)^2
\mytag{4.23}
$$
The formula \mythetag{4.23} means that the form $Q$ is positive, i.\,e\. 
$Q(\bold c,\bold c)\geqslant 0$. In most practical cases this form is strongly
positive, i.\,e\. $Q(\bold c,\bold c)>0$ for all $\bold c\neq 0$.\par
     It is well known (see \mycite{14}) that any quadratic form $Q$ in a linear 
Euclidean space $\Bbb E$ is associated with some symmetric operator 
$Q\!:\,\Bbb E\to\Bbb E$ such that 
$$
\hskip -2em
Q(\bold x,\bold y)=(Q\bold x,\bold y)=(\bold x,Q\bold y).
\mytag{4.24}
$$
In our case $\dim\Bbb E=3$. Therefore the operator $Q$ in \mythetag{4.24} has three
eigenvalues $\lambda_1$, $\lambda_2$, $\lambda_3$ associated with three mutually
perpendicular eigenvectors $\bold e_1$, $\bold e_2$, $\bold e_3$ of the unit length:
$|\bold e_1|=|\bold e_2|=|\bold e_3|=1$ and $\bold e_i\perp\bold e_j$ for $i\neq j$.
Due to \mythetag{4.23} the eigenvalues $\lambda_1$, $\lambda_2$, $\lambda_3$ are 
non-negative. Without loss of generality we enumerate these eigenvalues in the
non-decreasing order: 
$$
\hskip -2em
0\leqslant\lambda_1\leqslant\lambda_2\leqslant\lambda_3.
\mytag{4.25}
$$
Using the eigenvalues \mythetag{4.25}, we define the following four cases:
\roster
\item"1)" the non-degenerate case, where $0<\lambda_1\leqslant\lambda_2
\leqslant\lambda_3$;
\item"2)" the simple degenerate case, where $0=\lambda_1<\lambda_2\leqslant\lambda_3$;
\item"3)" the double degenerate case, where $0=\lambda_1=\lambda_2\leqslant\lambda_3$;
\item"4)" the triple degenerate case, where $0=\lambda_1=\lambda_2=\lambda_3$.
\endroster\par
     Our next goal is to find the minimum of the expression in the right hand side of
\mythetag{4.21} with respect to the variable $\bold c$. It is well-known that minima
(as well as other extrema) of multivariate functions are defined by means of 
the gradient equation:
$$
\hskip -2em
\nabla_{\bold c}(\bar D^{\kern 1.3pt 2})=\grad_{\,\bold c}(\bar D^{\kern 1.3pt 2})=0.
\mytag{4.26}
$$
Applying \mythetag{4.26} to \mythetag{4.21}, we derive the operator equation
$$
\hskip -2em
2\,Q\,\bold c-\bold L=0.
\mytag{4.27}
$$  
In the {\bf non-degenerate case} the operator equation \mythetag{4.27} is easily 
solvable:
$$
\hskip -2em
\bold c=\frac{1}{2}\,Q^{-1}\,\bold L=\frac{(\bold e_1,\bold L)\,\bold e_1}
{2\,\lambda_1}+\frac{(\bold e_2,\bold L)\,\bold e_2}
{2\,\lambda_2}+\frac{(\bold e_3,\bold L)\,\bold e_3}
{2\,\lambda_3}.
\mytag{4.28}
$$
The solution \mythetag{4.28} of the equation \mythetag{4.27} is unique. It corresponds 
to the minimum of the expression in the right hand side of \mythetag{4.21}. Substituting
\mythetag{4.28} back into \mythetag{4.21}, we derive the following formula for the minimal value 
of $\bar D^{\kern 1.3pt 2}$:
$$
\hskip -2em
\bar D^{\kern 1.3pt 2}=M-\frac{(\bold e_1,\bold L)^2}
{\lambda_1}-\frac{(\bold e_2,\bold L)^2}{\lambda_2}
-\frac{(\bold e_3,\bold L)^3}{\lambda_3}.
\mytag{4.29}
$$\par
     In the {\bf degenerate cases} the formula \mythetag{4.29} is not applicable. 
The degenerate cases are served by the following lemmas. 
\mylemma{4.2} The eigenvalue $\lambda_k$ of the operator $Q$ in \mythetag{4.24} 
vanishes if and only if all of the points $X_1,\,\ldots,\,X_n$ belong to the plane 
given by the equation 
$$
\hskip -2em
(\bold r-\bold r_{\text{cm}},\bold e_k)=0,
\mytag{4.30}
$$
where the radius-vector of the center of mass $\bold r_{\text{cm}}$ is given by the 
formula \mythetag{4.22}. 
\endproclaim
\mylemma{4.3} If the eigenvalue $\lambda_k$ of the operator $Q$ in \mythetag{4.24}
is equal to zero, then the vector $\bold L$ given by the formula \mythetag{4.19} is
perpendicular to the corresponding eigenvector $\bold e_k$, i\.\,e\. 
$(\bold L,\bold e_k)=0$. 
\endproclaim
\demo{Proof of Lemma~\mythelemma{4.2}} Due to \mythetag{4.24} the equality $\lambda_k=0$
implies $Q(\bold e_k,\bold e_k)=0$. Therefore, applying the formula \mythetag{4.23}, we 
derive
$$
\hskip -2em
\sum^n_{i=1}(\bold r_i-\bold r_{\text{cm}},\bold e_k)^2=0.
\mytag{4.31}
$$
The sum of the squares of real numbers is equal to zero if and only if each number is
equal to zero. Hence \mythetag{4.31} implies $(\bold r_i-\bold r_{\text{cm}},\bold e_k)=0$
for all $i=1,\,\ldots,\,n$, i.\,e. the the radius vector of each point $\bold r=\bold r_i$ 
in the group $X_1,\,\ldots,\,X_n$ obeys the equation \mythetag{4.30}. Lemma~\mythelemma{4.2}
is proved.\qed\enddemo
\demo{Proof of Lemma~\mythelemma{4.3}} Let's calculate the scalar product
$(\bold L,\bold e_k)$ using \mythetag{4.19}: 
$$
(\bold L,\bold e_k)=\frac{1}{n}\sum^n_{i=1}(\bold r_i,\bold e_k)
\,|[\bold r_i,\bold a]|^2-\biggl(\frac{1}{n}\sum^n_{i=1}(\bold r_i,\bold e_k)\biggr)
\biggl(\frac{1}{n}\sum^n_{i=1}|[\bold r_i,\bold a]|^2\biggr).
\quad
\mytag{4.32}
$$
Applying \mythetag{4.22} to the formula \mythetag{4.32}, we bring it to 
$$
\hskip -2em
\gathered
(\bold L,\bold e_k)=\frac{1}{n}\sum^n_{i=1}(\bold r_i,\bold e_k)
\,|[\bold r_i,\bold a]|^2-(\bold r_{\text{cm}},\bold e_k)
\biggl(\frac{1}{n}\sum^n_{i=1}|[\bold r_i,\bold a]|^2\biggr)=\\
=\frac{1}{n}\sum^n_{i=1}\bigl((\bold r_i,\bold e_k)
-(\bold r_{\text{cm}},\bold e_k)\bigr)\,|[\bold r_i,\bold a]|^2.
\endgathered
\mytag{4.33}
$$
Since $\lambda_k=0$, Lemma~\mythelemma{4.2}, which is already proved, yields
$(\bold r_i,\bold e_k)=(\bold r_{\text{cm}},\bold e_k)$. Hence the last sum
in \mythetag{4.33} is equal to zero, which implies $(\bold L,\bold e_k)=0$.
This means that Lemma~\mythelemma{4.3} is proved. 
\qed\enddemo
     Note that in the {\bf double degenerate case} Lemma~\mythelemma{4.2} says 
that the points $X_1,\,\ldots,\,X_n$ belong to the intersection of two
planes given by the equations
$$
\xalignat 2
&\hskip -2em
(\bold r-\bold r_{\text{cm}},\bold e_1)=0,
&&(\bold r-\bold r_{\text{cm}},\bold e_2)=0.
\mytag{4.34}
\endxalignat
$$
The intersection of the planes \mythetag{4.34} is the straight line $l$ passing through 
the center of mass point and directed along the vector $\bold e_3$. It is clear
that the solution of the best fitting cylinder problem in this case cannot be unique.
Indeed, any cylinder whose axis is parallel to the line $l$ and whose surface comprises 
the line $l$ is a best fitting cylinder for the group of points $X_1,\,\ldots,\,X_n$ 
belonging to the line $l$.\par 
     In the {\bf triple degenerate case}  Lemma~\mythelemma{4.2} says 
that the points $X_1,\,\ldots,\,X_n$ belong to the intersection of three 
planes given by the equations
$$
\xalignat 3
&(\bold r-\bold r_{\text{cm}},\bold e_1)=0,
&&(\bold r-\bold r_{\text{cm}},\bold e_2)=0,
&&(\bold r-\bold r_{\text{cm}},\bold e_3)=0.
\qquad
\mytag{4.35}
\endxalignat
$$
The intersection of the planes \mythetag{4.35} is a set consisting of exactly one point
which is the center of mass of the points $X_1,\,\ldots,\,X_n$. Therefore in this case
the points $X_1,\,\ldots,\,X_n$ do coincide with each other and with their center of mass.
Any cylinder whose surface comprises the center of mass point is a best fitting cylinder for them.\par
    In the {\bf simple degenerate case} the solution of the best fitting cylinder problem
can also be not unique. In this case Lemma~\mythelemma{4.2} says that the points $X_1,\,\ldots,\,X_n$ 
belong to the plane given by the equation
$$
\hskip -2em
(\bold r-\bold r_{\text{cm}},\bold e_1)=0.
\mytag{4.36}
$$
Let's imagine two parallel straight lines $l_1$ and $l_2$ on the plane \mythetag{4.36} 
and imagine a group of points $X_1,\,\ldots,\,X_n$ belonging to these two lines. 
\pagebreak There are infinitely many cylinders whose surfaces comprise two parallel 
straight lines $l_1$ and $l_2$. All of them are best fitting cylinders for the points
$X_1,\,\ldots,\,X_n$ in this case.\par
     Due to the non-uniqueness observed in the above examples we shall not consider the
degenerate cases at all. Below in Section 5 we complete the solution of the problem for
the non-degenerate case only which is the most practical one.\par
\head
5. The solution of the problem for the non-degenerate case. 
\endhead
     The expression \mythetag{4.29} for $\bar D^{\kern 1.3pt 2}$ comprises the scalar 
parameter $M$ and the vectorial parameter $\bold L$. They are given by the formulas 
\mythetag{4.20} and \mythetag{4.19} respectively. The parameter $M$ is quartic with 
respect to $\bold a$, while $\bold L$ is quadratic with respect to $\bold a$. Upon 
substituting \mythetag{4.20} and \mythetag{4.19} into \mythetag{4.29} we get an expression
for $\bar D^{\kern 1.3pt 2}$ which is quartic with respect to $\bold a$. The vector 
$\bold a$ can be presented through its expansion in the orthonormal basis of the 
eigenvectors $\bold e_1$, $\bold e_2$, $\bold e_3$ of the operator $Q$:
$$
\hskip -2em
\bold a=\sum^3_{i=1}a^i\,\bold e_i.
\mytag{5.1}
$$
We use upper indices $a^1$, $a^2$, $a^3$ for the components of the vector $\bold a$
in \mythetag{5.1} according to Einstein's tensorial notation (see \mycite{13}). In terms
of these components the quartic expression for $\bar D^{\kern 1.3pt 2}$ can be written as 
follows: 
$$
\hskip -2em
\bar D^{\kern 1.3pt 2}=\sum^3_{i=1}\sum^3_{j=1}\sum^3_{k=1}\sum^3_{q=1}
D_{ijkq}\,a^i\,a^j\,a^k\,a^q.
\mytag{5.2}
$$
Through $D_{ijkq}$ in \mythetag{5.2} we denote the components of a symmetric tensor
$\bold D$. Due to the symmetry not all of them are independent. Here is the list of 
independent components: $D_{1111}$, $D_{2222}$, $D_{3333}$, $D_{1112}$, $D_{1113}$, 
$D_{1222}$, $D_{2223}$, $D_{1333}$, $D_{2333}$, $D_{1122}$, $D_{1133}$, $D_{2233}$, 
$D_{1123}$, $D_{1223}$, $D_{1233}$. Using them, we can write \mythetag{5.2} as 
$$
\hskip -2em
\gathered
\bar D^{\kern 1.3pt 2}=D_{1111}\,(a^1)^4+D_{2222}\,(a^2)^4+D_{3333}\,(a^3)^4\,+\\
+\,4\,D_{1112}\,(a^1)^3\,a^2+4\,D_{1113}\,(a^1)^3\,a^3+4\,D_{1222}\,a^1\,(a^2)^3\,+\\
+\,4\,D_{2223}\,(a^2)^3\,a^3+4\,D_{1333}\,a^1\,(a^3)^3+4\,D_{2333}\,a^2\,(a^3)^3\,+\\
+\,6\,D_{1122}\,(a^1)^2\,(a^2)^2+6\,D_{1133}\,(a^1)^2\,(a^3)^2
 +6\,D_{2233}\,(a^2)^2\,(a^3)^2\,+\\
+\,12\,D_{1123}\,(a^1)^2\,a^2\,a^3+12\,D_{1223}\,a^1\,(a^2)^2\,a^3
 +12\,D_{1233}\,a^1\,a^2\,(a^3)^2.
\endgathered
\mytag{5.3}
$$
The components $a^1$, $a^2$, $a^3$ of the vector $\bold a$ in \mythetag{5.1} are 
not independent. From \mythetag{4.2} we derive the following relationship for them:
$$
\hskip -2em
(a^1)^2+(a^2)^2+(a^3)^2=1.
\mytag{5.4}
$$
\par
    The problem now is reduced to finding the minimum of the expression \mythetag{5.3}
under the restriction \mythetag{5.4}. Taking $a^1$ and $a^2$ for independent 
variables and using \mythetag{5.4}, one can express $a^3$ as a function $a^3=a^3(a^1,a^2)$. 
Then 
$$
\xalignat 2
&\hskip -2em
\frac{\partial a^3}{\partial a^1}=-\frac{a^1}{a^3},
&&\frac{\partial a^3}{\partial a^2}=-\frac{a^2}{a^3}.
\mytag{5.5}
\endxalignat
$$
The minimum of \mythetag{5.3} is determined by the following two equations:
$$
\xalignat 2
&\hskip -2em
\frac{\partial\bar D^{\kern 1.3pt 2}}{\partial a^1}=0,
&&\frac{\partial\bar D^{\kern 1.3pt 2}}{\partial a^2}=0.
\mytag{5.6}
\endxalignat
$$
When calculating the derivatives \mythetag{5.6} we take into account \mythetag{5.5}.
As a result we derive two homogeneous polynomial equations of the fourth degree:
$$
\xalignat 2
&\hskip -2em
P_{31}(a^1,a^2,a^3)=0,
&&P_{32}(a^1,a^2,a^3)=0.
\mytag{5.7}
\endxalignat
$$\par
     Taking $a^2$ and $a^3$ for independent variables and acting in a similar way as above,
we derive the other two homogeneous polynomial equations of the fourth degree
$$
\xalignat 2
&\hskip -2em
P_{12}(a^1,a^2,a^3)=0,
&&P_{13}(a^1,a^2,a^3)=0.
\mytag{5.8}
\endxalignat
$$
Then we take $a^3$ and $a^1$ for independent variables and repeat the procedure. As a result
we obtain two more homogeneous polynomial equations of the fourth degree:
$$
\xalignat 2
&\hskip -2em
P_{23}(a^1,a^2,a^3)=0,
&&P_{21}(a^1,a^2,a^3)=0.
\mytag{5.9}
\endxalignat
$$
Not all of the equations \mythetag{5.7}, \mythetag{5.8}, and \mythetag{5.9} are independent,
e\.\,g\. we have the following relationships for the polynomials in them: 
$$
\align
&P_{12}(a^1,a^2,a^3)+P_{21}(a^1,a^2,a^3)=0,\\
&P_{23}(a^1,a^2,a^3)+P_{32}(a^1,a^2,a^3)=0,\\
&P_{31}(a^1,a^2,a^3)+P_{13}(a^1,a^2,a^3)=0.
\endalign
$$
The explicit expressions for these polynomials are given in a machine-readable form in the ancillary 
file \darkred{polynomials.txt}.\par
     The equations \mythetag{5.7}, \mythetag{5.8}, \mythetag{5.9} along with the equation
\mythetag{5.4} are sufficient for finding a unit vector $\bold a$ corresponding to the minimum
of $\bar D^{\kern 1.3pt 2}$ in \mythetag{5.3}. This minimum does exist due to the Weierstrass's
extreme value theorem (see \mycite{15}) since $\bar D^{\kern 1.3pt 2}$ in \mythetag{5.3} is a continuous 
function on the unit sphere \mythetag{5.4}, which is a compact set. We say that the solution given by 
the equations \mythetag{5.7}, \mythetag{5.8}, \mythetag{5.9}, and
\mythetag{5.4} is {\bf almost analytic} since it is not given by explicit formulas.\par  
\head
6. Summary and conclusions. 
\endhead
      Once the vector $\bold a$ solving the equations \mythetag{5.7}, \mythetag{5.8}, 
\mythetag{5.9}, and \mythetag{5.4} and providing the minimal value for 
$\bar D^{\kern 1.3pt 2}$ in \mythetag{5.3} is found (either numerically or by means of
symbolic computations), we apply it in order to calculate $\bold c$ in \mythetag{4.28}.
Then we substitute $\bold c$ into \mythetag{4.16} and calculate $\bold b$. Using $\bold a$
and $\bold b$, we calculate $\rho_i$ in \mythetag{4.3}. And finally, we substitute
$\rho_i$ into \mythetag{4.7} and calculate the radius $\rho$ of the cylinder best fitting 
the points $X_1,\,\ldots,\,X_n$ which were initially given.\par
      The above solution of the cylindrical regression problem is a little bit more close 
a {\bf completely analytic} solution than the original one in \mycite{12}. We expect that
some day, using prospective computers and packages for symbolic computations, one would
be able to resolve the equations \mythetag{5.7}, \mythetag{5.8}, \mythetag{5.9}, and \mythetag{5.4} 
analytically in their general form, \pagebreak i\.\,e\. keeping symbolic coefficients 
$D_{1111}$, $D_{2222}$, $D_{3333}$, $D_{1112}$, $D_{1113}$, 
$D_{1222}$, $D_{2223}$, $D_{1333}$, $D_{2333}$, $D_{1122}$, $D_{1133}$, $D_{2233}$, 
$D_{1123}$, $D_{1223}$, $D_{1233}$ in \mythetag{5.3}. But even in this case it's quite
likely that the solution obtained would be very huge and not human-observable.\par

\enddocument
\end